# Deterministic MDPs with Adversarial Rewards and Bandit Feedback

**Raman Arora**
TTIC
6045 S. Kenwood Ave.
Chicago, IL 60637, USA

**Ofer Dekel**
Microsoft Research
1 Microsoft Way
Redmond, WA 98052, USA

**Ambuj Tewari**
Department of Statistics
University of Michigan
Ann Arbor, MI 48109, USA

## Abstract

We consider a Markov decision process with deterministic state transition dynamics, adversarially generated rewards that change arbitrarily from round to round, and a bandit feedback model in which the decision maker only observes the rewards it receives. In this setting, we present a novel and efficient online decision making algorithm named *MarcoPolo*. Under mild assumptions on the structure of the transition dynamics, we prove that MarcoPolo enjoys a regret of $O(T^{3/4}\sqrt{\log T})$ against the best deterministic policy in hindsight. Specifically, our analysis does not rely on the stringent unichain assumption, which dominates much of the previous work on this topic.

## 1 INTRODUCTION

Sequential decision making problems come in a variety of different flavors. In some cases it is natural to model the environment as a stochastic process, while in other cases we require the more robust assumption that the environment behaves adversarially. In both cases, we can either assume that the environment is stationary or that it evolves over time. For some problems it suffices to assume that the decision-making agent is stateless, while for other problems we may need to assume that the agent has a state that depends on its past actions. We must also specify what level of feedback the agent observes after making each decision. Finally, if we assume that the agent indeed has a state, we must decide whether the state transition dynamics are deterministic or stochastic, and whether they are known or unknown to the agent.

In this paper, we consider sequential decision making in an adversarial and nonstationary environment where the agent has a state. That is, on each round of the sequential decision-making problem, the agent is in one of finitely many states $\mathcal{S}$ and performs an action from a finite action set $\mathcal{A}$. The rewards received by the agent depend both on its action and its state, and the mapping from state-action pairs to rewards is controlled by an adversary and changes arbitrarily over time. We assume that the only feedback that the agent observes is the value of the rewards that it receives, and specifically, that it does not get to see the rewards that it could have received if it had acted differently or if it were in a different state. This limited type of feedback is commonly known as *bandit feedback*. In fact, this feedback model is so natural that a typical researcher in reinforcement learning may not even add the qualifier "bandit" to it. However, we prefer to explicitly mention it to distinguish our setting from early work (Even-Dar et al., 2009) on adversarial MDPs that assumed more extensive feedback.

We also assume that the state transition dynamics are deterministic and fully known to the agent. However, we note that, in the case of deterministic dynamics, it is easy to relax the assumption that the dynamics are known. Following ideas of Ortner (2010), the agent can spend poly($|\mathcal{S}|, |\mathcal{A}|$) time figuring out the dynamics. Therefore, we assume that the agent knows its current state and knows which state it will transition to as a result of each action. This type of state transition model is often called a *deterministic Markov decision process* (DMDP).

Our adversarial reward setting makes sense when the agent faces other agents with conflicting interests. This happens, for instance, when robots interact with other robots, when bidding agents compete against each other in auctions, when trading agents compete to make money, or when the agent is a character in a video game. In these strategic and adversarial settings, the environment is likely to change in unpredictable and arbitrary ways. In all of the aforementioned decision problems, bandit feedback seems to be the most realistic feedback model. Although we make pessimistic assumptions about the world, we emphasize that our assumptions are strictly more general than the alternative: if the agent is lucky enough to find itself in a stochastic or stationary environment, or if richer feedback is available, our setting still applies.

The harsh assumptions about the rewards and the limited feedback are contrasted by the optimistic assumption that the state transition dynamics form a DMDP. In some cases, deterministic state transitions are a realistic assumption while in other cases this assumption may be overly simplistic. A stochastic system is often well approximated by a deterministic system (Bertsekas, 2005, Chapter 2). For instance, consider the motivating example in Even-Dar et al. (2009), where the goal is to drive along the shortest path from point $a$ to point $b$. Although driving a car involves stochastic uncertainties, it is usually reasonable to assume that we have deterministic control over which streets our car takes along the path to our destination. Additionally, many of the uncertainties of driving a car can be modeled as part of the adversarial rewards. Another situation that leads to modeling using DMDPs is when the adversary is a finite state machine whose transitions depend on (equivalence classes of) the history of the agent's actions (Maillard and Munos, 2011).

Modeling the state transitions as a DMDP enables us to handle the adversarial bandit-feedback setting in a principled and rigorous fashion. Obtaining a similar result in the general weakly communicating MDP setting (where state transitions are stochastic) remains an elusive open problem. We hope that the techniques developed in this paper will prove useful toward solving that problem.

Another important design choice is how to evaluate the agent's actions. For simplicity, we assume that every action is available at each state, and we note that this assumption can be easily relaxed. Since the rewards are controlled by an adversary, they only become semantically meaningful when we compare the agent's cumulative reward with a suitable benchmark. We let $\Pi$ be the set of all deterministic policies, where each $\pi \in \Pi$ is a mapping from $\mathcal{S}$ to $\mathcal{A}$, and we choose our benchmark to be the maximal reward obtained by any policy in $\Pi$. In other words, we compare the agent's rewards to those of the best deterministic policy in hindsight. Letting $r_1, r_2, \ldots$ denote the adversarial sequence of reward functions (chosen in advance by the adversary before the agent begins its moves), letting $(s_t^\pi, a_t^\pi)_{t=1}^\infty$ denote the trajectory of state-action pairs generated by policy $\pi$, and letting $(s_t, a_t)_{t=1}^\infty$ denote the trajectory of state-action pairs generated by the agent, we define the agent's undiscounted *regret* after $T$ rounds as

$$\max_{\pi \in \Pi} \sum_{t=1}^T r_t(s_t^\pi, a_t^\pi) - \mathbb{E}\left[\sum_{t=1}^T r_t(s_t, a_t)\right] \enspace .$$

The agent's rewards appear in expectation because we allow randomized decisions. We say that the agent is learning if its regret can be upper-bounded by a sub-linear function of $T$, uniformly for all sequences of reward functions.

As mentioned above, we allow the adversary to modify the rewards for any state-action pair without any restrictions. Therefore, the same sequence of state-action pairs may have high rewards at one time and low rewards at another time. It could very well be that a given action sequence (starting from a given state $s$) may produce different rewards depending on whether we start performing the sequence on an odd or even time step. Therefore, it is insufficient to merely discover high-reward paths through the state space; we must also consider timing issues.

The main contribution of this paper is *MarcoPolo*, an efficient algorithm that enjoys a regret bound of $O(T^{3/4}\sqrt{\log T})$ against the best deterministic policy in hindsight (Theorem 2). The name MarcoPolo is an approximate acronym for **M**DP with **A**dversarial **R**ewards, weakly **CO**mmunicating structure, **P**artial feedback, by reduction to **O**nline **L**inear **O**ptimization. Anecdotally, Marco Polo was a famous Venetian merchant traveler who explored the world in search of high rewards, just as our algorithm explores the DMDP state space in search of rewards. Moreover, Marco Polo is the name of a well-known children's game that is played in a swimming pool, where one swimmer has to capture the other swimmers using only bandit feedback.

The basic idea behind MarcoPolo is quite simple. We observe that a deterministic policy repeats the same cycle of actions again and again. We then realize that if an oracle were to tell us the cycle length of the best deterministic policy and some additional information about its trajectory, then we could do as well as the best action cycle via a reduction to the *bandit linear optimization* (BLO) setting. Such an oracle is unavailable to us, but we can do almost as well as the oracle using a *multi-armed bandit* (MAB) algorithm. The MAB algorithm explores different cycle lengths and trajectories by invoking the BLO algorithm multiple times. To the best of our knowledge, this two-tier hierarchy involving a MAB algorithm that invokes a BLO algorithm is a novel conceptual contribution to the design space of regret minimization algorithms.

## 1.1 RELATED WORK

Our setting has much in common with a line of research initiated by Even-Dar et al. (2009) and further developed by Yu et al. (2009) and Neu et al. (2010), which deals with MDPs with adversarial rewards. Even-Dar et al. (2009) assumes a full-information feedback model while Yu et al. (2009) and Neu et al. (2010) only require bandit feedback. While these papers assume that the state transition dynamics are stochastic, they also assume that these dynamics have a *unichain* structure. This means that *every* policy must lead to a Markov chain with a single recurrent class (plus some possibly empty set of transient states). This is a very restrictive assumption, as we discuss in Sec. 2.1. In our setting, we assume that the state transition dynamics are deterministic and *weakly communicating*, i.e. it is possible to move from any state to any other state under *some* policy. Although deterministic transition dynamics are less general than stochastic ones, our analysis requires a much weaker assumption on the connectivity between states.

The work in Yu and Mannor (2009) also deals with MDPs with adversarial rewards, bandit feedback, and stochastic state transition dynamics. Moreover, it allows the stochastic transition dynamics to change over time in an adversarial way. The analysis in Yu and Mannor (2009) relies on the strong assumption that the MDP has a finite mixing time. Again, our analysis has the advantage that it relies on a weaker assumption on state connectivity.

Ties between the reinforcement learning literature and the area of individual sequence prediction and compression in information theory are further developed in Farias et al. (2010).

There is a well developed regret minimization and PAC-MDP literature that deals with a similar problem, except that the rewards are stochastic and not adversarial. We cannot do justice to that extensive literature but merely refer the reader to the survey by Szepesvari (2010) and the references therein. Our work bears some similarities to the work of Ortner (2010), which considers sequential decision making in DMDPs with bandit feedback. However, the rewards in that paper are stochastic and not adversarial, hence, the motivation and techniques used are rather different.

The importance of developing machine learning algorithms that work in reactive environments (where agent's actions impact the future evolution of rewards) has caught the attention of many researchers. de Farias and Megiddo (2006) propose algorithms that combine expert advice in a reactive environment. Ryabko and Hutter (2008) consider the problem of learnability in a very general reactive setting and show that environments that allow a rapid recovery from mistakes are asymptotically learnable. Arora et al. (2012) introduce the notion of policy regret as a more meaningful alternative to regret for measuring an online algorithm's performance against adaptive adversaries. Arora et al. (2012) also presents a general reduction that converts any bandit algorithm with a sublinear regret into an algorithm with a sublinear policy regret.

## 2 PRELIMINARIES

Let $\mathcal{S}$ be a set of $n$ states and let $\mathcal{A}$ be a finite set of actions. An *adversarial-reward deterministic Markov decision process* (DMDP) is a tuple $(\mathcal{S}, \mathcal{A}, f, r_{1:T})$ where $\mathcal{S}$ is a finite set of *states*, $\mathcal{A}$ is a finite set of *actions*, $f : \mathcal{S} \times \mathcal{A} \to \mathcal{S}$ is a deterministic *state transition function*, which maps a state-action pair to the next state, and $r_{1:T}$ is a sequence $(r_t)_{t=1}^T$ of *reward functions* $r_t : \mathcal{S} \times \mathcal{A} \to [0, 1]$.

We slightly overload our notation and allow the second argument of the state transition function to be a sequence of actions, rather than a single action. Namely, $f(s, (a_1, \ldots, a_k))$ is the state we reach if we start from $s$ and perform the action sequence $a_1, \ldots, a_k$. More formally, the extended definition of $f$ is given by the recursion

$$f\big(s, (a_1, \ldots, a_k)\big) \;=\; f\Big(f\big(s, (a_1, \ldots, a_{k-1})\big),\; a_k\Big) \;\;.$$

We also find it useful to denote the set of state-action pairs that lead to a given state $s$ by $\mathrm{I}(s)$. That is,

$$\mathrm{I}(s) = \{(s', a') \,:\, f(s', a') = s\} \;\;.$$

A deterministic *policy* is a mapping $\pi : \mathcal{S} \to \mathcal{A}$. Together with an initial state $s_1$, each policy defines a sequence of actions and states. Since $\mathcal{S}$ is finite, a state will eventually repeat, and from then onwards the policy will keep repeating the same cycle of actions. We categorize a policy $\pi$ by the length of this cycle and we let $\Pi_k$ denote the set of deterministic policies that induce cycles of length $k$. We prove a regret bound that compares the algorithm's cumulative reward with the reward of the best deterministic policy in $\Pi_{\leq L} = \bigcup_{k \leq L} \Pi_k$, where $L$ is a user-defined constant. Since the cycle length is bounded by $n$, setting $L = n$ gives a regret bound that compares the algorithm to all deterministic policies. The benefit of proving a regret bound for $L < n$ is to demonstrate how the bound degrades gracefully with the complexity of the competitor class.

Our analysis deals with weakly communicating DMDPs (see Section 2.1 for definitions of weakly communicating MDP and the associated closed set of states) and requires the following additional assumption.

**Assumption 1.** *For any pair of states $s \neq s'$ in the closed set, there exists an action sequence $a = (a_1, \ldots, a_d)$ (of length exactly $d$) such that $f(s, a) = s'$.*

Note that the assumption only talks about *existence* of an action sequence. We note that finding the action sequence can be done efficiently, e.g. using dynamic programming. This assumption is a very mild one and it may well be possible to dispense with it entirely. A simple sufficient condition that implies this assumption is the existence of at least one self-loop in the DMDP. Another sufficient condition is that the transition graph is symmetric (i.e. if $f(s, a) = s'$ then there exists an opposite action $a' \in \mathcal{A}$ such that $f(s', a') = s$) and there exists an odd length cycle in the graph.

### 2.1 THE UNICHAIN CONDITION IS TOO STRINGENT

The difference between a DMDP and a general MDP is that, in the latter, the transition dynamics are stochastic, i.e. $f$ maps an $(s, a)$ pair to a *distribution* over states. Two well known subclasses of MDPs are *recurrent* (or ergodic) and *communicating* MDPs. In a recurrent MDP, *every* policy $\pi$ induces a Markov chain with a single recurrent class. The more general notion of a communicating MDP only requires that it be possible to move (with non-zero probability) from any state to any other state using *some* policy. Note the difference in quantifiers — "every" versus "some" — in the definitions of recurrent and communicating MDPs. Thus, the recurrent condition is much more stringent than the communicating condition.

Both notions can be generalized a little bit to allow for some transient states. Thus, we arrive at the classes of *unichain* and *weakly communicating* MDPs. In unichain MDPs, *every* policy induces a Markov chain with a single recurrent class plus a possibly empty set of transient states. In weakly communicating MDPs, there is a *closed* set of states such that each state is accessible, under *some* policy, from any other state in the closed set (with non-zero probability under some policy) plus a possibly empty set of states that are transient under every policy. In the same way as recurrent is a much stronger condition compared to communicating, the unichain condition is much stronger compared to weakly communicating.

We have the (strict) inclusions:

recurrent $\subset$ unichain $\subset$ weakly communicating ,
recurrent $\subset$ communicating $\subset$ weakly communicating .

The graph associated with a DMDP is a directed graph with $\mathcal{S}$ as the set of vertices and the set of edges given by

$$\{(s, s') \ : \ \exists a \in \mathcal{A} \text{ s.t. } f(s, a) = s'\} \ .$$

For DMDPs, the unichain condition can be equivalently stated in terms of its graph. A DMDP is unichain iff *any* two cycles in its graph have a common vertex (Feinberg and Yang, 2008). A DMDP is weakly communicating if, after removing a possibly empty set of states that are transient under all policies, we get a graph that is strongly connected. Thus, any learning algorithm will, after at most $n$ steps, end up in the strongly connected component. Therefore, in our regret analysis, we will assume that the DMDP is communicating.

The "no vertex-disjoint cycles" characterization of the unichain condition for DMDPs shows again why the unichain condition is so stringent. Two self-loops on two distinct states will violate it. It is also not preserved under very simple operations of building larger MDPs from smaller ones. For instance, suppose we start from two unichain DMDPs on disjoint state spaces $S_1, S_2$ and join them by introducing two actions $a_1, a_2$ such that $f(s_1, a_1) = s_2$ and $f(s_2, a_2) = s_1$ for some $s_1 \in S_1, s_2 \in S_2$, then the resulting DMDP will not, in general, be unichain (but will be weakly communicating).

## 3 SOLUTION FOR FIXED LENGTH AND STATE

As a warm-up, we consider a toy version of our problem, where the competitor class is restricted to a smaller subset of $\Pi_{\leq L}$. Assume that we are told to focus on a specific cycle length $k$ and a specific initial state $\bar{s}$. Let $\Pi_{k,\bar{s}}$ denote the subset of policies in $\Pi_k$ that start in state $\bar{s}$ at time 1 and return to $\bar{s}$ at time $k + 1$. Assume that we are only required to compete with policies in $\Pi_{k,\bar{s}}$.

Next, we characterize the action sequences induced by policies in $\Pi_{k,\bar{s}}$. Define the set

$$\mathcal{C}_{k,\bar{s}} \ = \ \{c \in \mathcal{A}^k \ : \ f(\bar{s}, c) = \bar{s}\} \ .$$

$\mathcal{C}_{k,\bar{s}}$ is the set of $k$-length action sequences such that starting from state $\bar{s}$ and performing an action sequence in $\mathcal{C}_{k,\bar{s}}$ brings us back to state $\bar{s}$. We say that $\mathcal{C}_{k,\bar{s}}$ is the set of *length $k$ action cycles* with respect to a starting state $\bar{s}$. Note that this set contains all of the simple cycles of length $k$, but may also contain cycles that pass through $\bar{s}$. In this section, we present a randomized online algorithm that competes with any action sequence that consists of recurrences of some cycle in $\mathcal{C}_{k,\bar{s}}$. Note that this set of competitors may also contain action sequences that are not induced by any policy in $\Pi_{k,\bar{s}}$, for example, due to the fact that $\mathcal{C}_{k,\bar{s}}$ is not restricted to simple cycles.

First, we present a general outline of our algorithm. There is an initial "lock-in" part in the algorithm to take care of the fact that the algorithm will eventually be called as a subroutine when the global time clock reads $t_1$ units and the agent is in some arbitrary state $s_1 \in \mathcal{S}$. Thus, there is a "phase lock-in" problem: the competitors in $\mathcal{C}_{k,\bar{s}}$ start from $\bar{s}$ at global time 1 and return to it at times $k+1, 2k+1, \ldots$. The agent needs to move from $s_1$ and get "locked into" one of the cycles so that it is not out of phase. Here is where Assumption 1 comes in handy. The algorithm picks an arbitrary cycle $(c_1, \ldots, c_k) \in \mathcal{C}_{k,\bar{s}}$. At time $t_1 + d$, this competitor will be at state $c_{k'}$ where $k' = (t_1 + d - 1) \mod k + 1$. The algorithm performs a sequence of $d$ actions that bring it to $c_{k'}$ at time $t_1 + d$. Then we waste $k'' = (k - k' + 1) \mod k$ more time steps to come back to $\bar{s}$ and the agent is now "in phase" with $\mathcal{C}_{k,\bar{s}}$.

The algorithm then splits the remaining rounds into epochs of length $k$, where epoch $j$ starts on round $d + k'' + (j-1)k + 1$ and ends on round $d + k'' + jk$. At the beginning of epoch $j$, the algorithm defines a distribution $\theta_j$ over the set $\mathcal{C}_{k,s}$ and samples an action sequence from this distribution. The algorithm then spends the next $k$ rounds executing the actions in the sequence that it has chosen. The algorithm sticks to the chosen sequence for the duration of the epoch, disregarding the rewards it observes along the way. Only at the end of the epoch does the algorithm look back on its rewards and defines a new distribution $\theta_{j+1}$ for the next epoch. In fact, our algorithm will only use the sum of the $k$ rewards, rather than the $k$ individual reward values.

It remains to specify how the algorithm sets the distribution $\theta_j$. It does so via a reduction of our problem to a *bandit linear optimization* (BLO) problem[1]. We briefly describe the BLO problem setting and the interested reader is referred to (Flaxman et al., 2005; Abernethy et al., 2008) for details. A BLO problem is a repeated game between a randomized algorithm and an adversary, which takes place over a predefined polyhedral set $\mathcal{X} \subset \mathbb{R}^n$. Before the game begins, the adversary secretly picks a sequence of vectors $\rho_1, \rho_2, \ldots$, each in $\mathbb{R}^n$. On round $i$ of the game, the algorithm begins by choosing a distribution over $\mathcal{X}$ and sampling a point $x_i$ from that distribution. Next, the adversary

---

[1]A similar approach is used in (Flaxman et al., 2005; Abernethy et al., 2008) to solve the *bandit shortest path* problem. For more information on the bandit shortest path problem, see (McMahan and Blum, 2004; Awerbuch and Kleinberg, 2004; György et al., 2007)

---

**Algorithm 1**: Fixed length $k$ and state $\bar{s}$

**input:** initial state $s_1$, global time $t_1$, length $k$, state $\bar{s}$, iterations $T$
**initialize:** BLO with $n = nk|\mathcal{A}|$ and $\mathcal{X}$ as defined in Eqs. (2 – 5)
Pick an arbitrary cycle $(c_1 = \bar{s}, \ldots, c_k) \in \mathcal{C}_{k,\bar{s}}$
Set $k' \leftarrow (t_1 + d - 1) \mod k + 1$
$(a_1, \ldots, a_d) \leftarrow$ get action sequence from $s_1$ to $c_{k'}$
**for** $t = 1, \ldots, d$
  perform action $a_t$: $s_{t+1} \leftarrow f(s_t, a_t)$
  observe (bandit) reward $r_t(s_t, a_t)$
  accumulate $R \leftarrow R + r_t(s_t, a_t)$
Set $k'' \leftarrow (k - k' + 1) \mod k$
Spend $k''$ time steps visiting the states $c_{k'+1}, \ldots, c_k$
Accumulate the $k''$ rewards in $R$
**for** $j = 1, 2, \ldots$
  $x_j \leftarrow$ get vector from BLO
  $\theta_j \leftarrow$ decompose$(x_j)$
  sample action cycle $(c_1 = \bar{s}, \ldots, c_k) \sim \theta_j$
  **for** $i = 1, \ldots, k$
    $t \leftarrow d + k'' + (j-1)k + i$
    **if** $t > T$ **then**
      **return** $(R, s_t)$
    perform action $a_t = c_i$: $s_{t+1} \leftarrow f(s_t, c_i)$
    observe (bandit) reward $r_t(s_t, c_i)$
    accumulate $R_j \leftarrow R_j + r_t(s_t, c_i)$
  provide $R_j$ as feedback to BLO
  $R \leftarrow R + R_j$

---

computes $\rho_i \cdot x_i$ and reveals the result to the algorithm, while still hiding the vector $\rho_i$. The value $\rho_i \cdot x_i$ is the algorithm's reward on round $i$, and cumulative reward over a sequence of $m$ rounds equals $\sum_{i=1}^{m} \rho_i \cdot x_i$. The goal is to accumulate a large reward over time. In this setting, the BLO algorithm presented in Abernethy et al. (2008) guarantees that the algorithm's expected regret after $m$ iterations is bounded as

$$\max_{x \in \mathcal{X}} \sum_{i=1}^{m} \rho_i \cdot x - \mathbb{E}\left[\sum_{i=1}^{m} \rho_i \cdot x_i\right] \leq 4Un^{3/2}\sqrt{m \log(m)} \quad , \tag{1}$$

where $U$ is an upper bound on $\rho_i \cdot x_i$ for all $i$.

A useful extension to the BLO setting is the double randomization technique, which involves adding an additional layer of randomization to the algorithm's strategy. The algorithm chooses the point $x_i$ normally, but then replaces $x_i$ with an *unbiased estimator* of $x_i$. In other words, the algorithm can choose any distribution $\theta_i$ over $\mathcal{X}$ such that $\mathbb{E}_{X \sim \theta_i}[X] = x_i$. The algorithm then samples a point $X_i$ according to $\theta_i$ and plays $X_i$ instead of $x_i$. Due to linearity, the expected reward $\mathbb{E}[\rho_i \cdot X_i | x_i]$ still equals $\rho_i \cdot x_i$, and the regret bound stated above remains valid. We use this exten-sion in our reduction below. For more details, please see Abernethy et al. (2008, Section 7).

We are now ready to define the reduction from our problem to the BLO problem. The BLO problem takes place in the space $\mathbb{R}^n$, where $n = n|\mathcal{A}|k$. In this space, each coordinate corresponds to a triplet $(s, a, i)$, where $s$ is a state in $\mathcal{S}$, $a$ is an action in $\mathcal{A}$, and $i$ is a phase index in $[k]$. For any vector $v \in \mathbb{R}^n$, we use the notation $v_{(s,a,i)}$ to refer to the element of $v$ that corresponds to the triplet $(s, a, i)$.

We begin by embedding the sequence of reward functions that occur during epoch $j$ in $\mathbb{R}^n$. The relevant reward functions are $r_{d+k''+jk-k+i}, \ldots, r_{d+k''+jk}$. We represent their values by a vector $\rho_j \in \mathbb{R}^n$ defined as

$$\forall s \in \mathcal{S}, a \in \mathcal{A}, i \in [k] \quad \rho_{j,(s,a,i)} = r_{d+k''+jk-k+i}(s, a) \ .$$

Next, we represent each cycle $c \in \mathcal{C}_{s,k}$ by a binary vector $x(c) \in \mathbb{R}^n$ defined as

$$x(c)_{s,a,i} = \begin{cases} 1 & \text{if } a = c_i \text{ and } s = f(\bar{s}, (a_1, \ldots, a_{i-1})) \\ 0 & \text{otherwise} \end{cases},$$

for all $s \in \mathcal{S}, a \in \mathcal{A}$, and $i \in [k]$. Note that $x(c)$ has exactly $k$ non-zero entries that indicate the state-action pairs encountered along the cycle $c$. Moreover, note that the cumulative reward collected while performing the action sequence $c$ on epoch $j$ simply equals the dot product $\rho_j \cdot x(c)$.

Let $x(\mathcal{C}_{k,\bar{s}}) = \{x(c)\}_{c \in \mathcal{C}_{k,\bar{s}}}$ denote the set of vectors induced by all of the action cycles in $\mathcal{C}_{k,\bar{s}}$. Selecting a sequence of cycles $c_1, c_2, \ldots$ from the set $\mathcal{C}_{k,\bar{s}}$ so as to maximize the reward defined by the adversarial sequence $r_1, r_2, \ldots$ is equivalent to the problem of selecting a sequence of points $x_1, x_2, \ldots$ from the set $x(\mathcal{C}_{k,\bar{s}}) \subset \mathbb{R}^n$ so as to maximize the linear rewards defined by the adversarial sequence $\rho_1, \rho_2, \ldots$ However, the latter is not quite a BLO problem yet, since the set of actions $x(\mathcal{C}_{k,\bar{s}})$ is not a polyhedral set.

To resolve this problem, we define the set of actions for the BLO algorithm to be $\mathcal{X} = \text{conv}(x(\mathcal{C}_{k,\bar{s}}))$, the convex hull of the set $x(\mathcal{C}_{k,\bar{s}})$. Since $x(\mathcal{C}_{k,\bar{s}})$ is a finite set, its convex hull is polyhedral. Now, we can use a BLO algorithm to generate a sequence of points in $\mathcal{X}$. The set $\mathcal{X}$ can be defined more concisely using network flow constraints (Bazaraa et al., 2010). In our case, $\mathcal{X}$ can be written as

$$x \geq 0, \quad \sum_{a \in \mathcal{A}} x_{(\bar{s},a,1)} = 1 \ , \tag{2}$$

$$\forall s \notin \mathcal{S} \setminus \{\bar{s}\}, \ a \in \mathcal{A}, \ x_{(s,a,1)} = 0 \ , \tag{3}$$

$$\forall (s', a') \notin \text{I}(\bar{s}) \ x_{(s',a',k)} = 0 \ , \tag{4}$$

$$\forall s \in \mathcal{S}, \ 2 \leq i \leq k,$$

$$\sum_{(s',a') \in \text{I}(s)} x_{(s',a',i-1)} - \sum_{a \in \mathcal{A}} x_{(s,a,i)} = 0 \ . \tag{5}$$

Equation (2) ensures convexity: all paths have non-negative weight and the total weight of all paths equals one. The remaining constraints ensure that every vector in the set is indeed a combination of valid cycles that start and end at $\bar{s}$. Specifically, Eqs. (3) and (4) require all paths to begin and end at $\bar{s}$, whereas (5) is the classic flow conservation constraint, which requires that the total weight of paths that enter each state be equal to the total weight that exits that state, at each phase $i$. This concise representation enables us to run the BLO algorithm efficiently.

However, now the BLO algorithm picks points in $\mathcal{X}$ that do not directly correspond to individual cycles in $\mathcal{C}_{k,\bar{s}}$. Instead, these points are convex combinations of points that correspond to individual cycles. This problem is easily resolved using the double randomization technique described above. Namely, we can find an unbiased estimator of the point chosen by the BLO algorithm that always chooses a vector in the discrete set $x(\mathcal{C}_{k,\bar{s}})$. We apply the classic theorem of Carathéodory (1911), which states that any point in the polyhedral set $\mathcal{X} \subset \mathbb{R}^n$ can be represented as a convex combination of at most $n+1$ extreme points (vertices) of that set. The extreme points of $\mathcal{X}$ are points in the set $x(\mathcal{C}_{k,\bar{s}})$ and correspond to individual cycles. Moreover, the decomposition of a point in $\mathcal{X}$ as a convex combination of $n+1$ points in $x(\mathcal{C}_{k,\bar{s}})$ can be efficiently computed using a greedy augmenting-path-style algorithm (Bazaraa et al., 2010, Thm. 2.1). In summary, we use the BLO algorithm to choose a point in $\mathcal{X}$, we represent this point as a convex combination of points in $x(\mathcal{C}_{k,\bar{s}})$, we interpret this convex combination as a distribution over the respective action cycles in $\mathcal{C}_{k,\bar{s}}$, and we sample a concrete cycle from this distribution.

The pseudo-code of our algorithm is given in Algorithm 1. As mentioned above, the pseudo-code references three external procedures: the procedure that finds a path of length $d$ from $s_1$ to $\bar{s}$, the BLO algorithm, and the decomposition procedure that represents any point in $\mathcal{X}$ as a distribution over $n+1$ points in $x(\mathcal{C}_{k,\bar{s}})$.

Next, we state a regret bound for our algorithm.

**Theorem 1.** *Let $r_1, r_2, \ldots$ be an arbitrary sequence of reward functions, perhaps adversarially generated. Assume that we run Algorithm 1 for $T$ rounds with this sequence, with parameters $k \leq L$ and $\bar{s} \in \mathcal{S}$, and let $(s_t, a_t)_{t=1}^T$ be the resulting sequence of state-action pairs (see Alg. 1). On the other hand, let $\pi^\star$ be any deterministic policy in $\Pi_{k,\bar{s}}$ and let $(s_t^\star, a_t^\star)_{t=1}^T$ be the sequence of state-action pairs defined by this policy.*

*Then the regret is upper bounded by*

$$\sum_{t=1}^{T} r_t(s_t^\star, a_t^\star) - \mathbb{E}\left[\sum_{t=1}^{T} r_t(s_t, a_t)\right]$$
$$\leq 4L^2(n|\mathcal{A}|)^{3/2}\sqrt{T\log(T)} + (2L+d) \ .$$

*Proof.* Since the rewards on each round are bounded in $[0, 1]$, the regret on the first $d$ rounds is bounded by $d$. Then $k'' = (k - k' + 1) \mod k < k$ more rounds are wasted before coming back to $\bar{s}$. The regret on the remaining $T - d - k''$ rounds is bounded by applying the bound for BLO in Eq. (1). Specifically, in our case, we use Hölder's inequality to bound

$$\rho_i \cdot x_i \leq \|\rho_i\|_\infty \|x_i\|_1 \leq k \ ,$$

and therefore we can set $U$ in Eq. (1) to $k$. The dimensionality is $n|\mathcal{A}|k$. Finally, the number of BLO iterations equals the number of epochs of our algorithm, which is $m = \lfloor (T - d - k'')/k \rfloor \leq T/k$. Plugging these values into Eq. (1) gives the bound

$$4k(nk|\mathcal{A}|)^{3/2}\sqrt{T/k \log(T/k)} \ ,$$

which is upper bounded by

$$4L^2(n|\mathcal{A}|)^{3/2}\sqrt{T\log(T)} \ .$$

Finally, if $T - d - k''$ does not divide by $k$, we have a suffix of at most $k$ actions that incur an additional regret of at most $L$. $\square$

## 4 SOLUTION TO THE GENERAL PROBLEM

The result obtained in the previous section can be interpreted as follows. Let $\pi^\star$ be the optimal fixed policy in $\Pi_{\leq L}$, let $k^\star$ be the length of its cycle. Assume, for simplicity, that $\pi^\star$ enters its cycle straightaway and let $s^\star$ be its state at time $t = 1$. If an oracle were to reveal $k^\star$ and $s^\star$ to us, Algorithm 1 would solve the problem with a regret of $O(L^2(n|\mathcal{A}|)^{3/2}\sqrt{T\log(T)})$. Although we do not have access to such an oracle, we can exploit the fact that there are only $Ln$ possible values of $(k, s)$ that we need to explore.

In this section, we present the MarcoPolo algorithm, which solves the general problem defined in Sec. 1. It does so by splitting the online rounds into episodes of length $\tau$ ($\tau$ is specified later on) and running an instance of Alg. 1 on each episode. At the beginning of episode $j$, the algorithm chooses values $(k_j, s_j)$ and runs Alg. 1 with these parameters for $\tau$ rounds. Note that Alg. 1 starts from scratch on each episode, and does not retain information from previous episodes. Also note that Alg. 1 starts from the state that we happen to be in due to the previous episodes. However, the initial "lock-in" part in Alg. 1 makes sure we get locked in phase with policies in $\Pi_{k_j, s_j}$.

**Algorithm 2**: MarcoPolo
**input:** episode length $\tau$
**initialize:** $\hat{s}_1 = s_1, t = 1$
**for** $j = 1, 2, \ldots$
 $(k_j, s_j) \leftarrow$ get length/state from MAB
 $(R_j, \hat{s}_{j+1}) \leftarrow$ Alg. 1 with input $(\hat{s}_j, t, k_j, s_j, \tau)$
 provide $R_j$ as feedback to MAB
 $R \leftarrow R + R_j$
 $t \leftarrow t + \tau$

It remains to specify how our algorithm chooses $(k_j, s_j)$ on each episode. Again, we resort to a reduction, this time to the *multi-armed bandit* problem (MAB). The MAB setting is very similar to the BLO setting described above, except that the algorithm's action set is the set of standard unit vectors $\mathcal{X} = \{e_1, \ldots, e_n\}$, rather than a polyhedral set. Namely, the adversary chooses a sequence of reward vectors $\rho_1, \rho_2, \ldots$ before the game begins. On iteration $i$, the algorithm chooses a standard unit vector $x_i$, which has a single non-zero coordinate, and its reward $\rho_i \cdot x_i$ is simply the value of $\rho_i$ at that coordinate. The algorithm never sees the full reward vector $\rho_j$.

The EXP3 algorithm (Auer et al., 2002) solves the MAB problem, and comes with the following regret bound: For any number of iterations $m$, the expected regret after $m$ iterations is bounded by

$$\max_{x \in \mathcal{X}} \sum_{i=1}^{m} \rho_i \cdot x - \mathbb{E}\left[\sum_{i=1}^{m} \rho_i \cdot x_i\right] \leq U\sqrt{7mn\log(n)} \ , \quad (6)$$

where $U$ is an upper bound on $\rho_i \cdot x_i$ for all $i$.

To simplify the presentation and analysis of the reduction to the MAB problem, we slightly modify the algorithm described above. Instead of choosing a single setting $(k_j, s_j)$ on episode $j$ and running Alg. 1 only on that setting, imagine that we run $nL$ copies of Alg. 1 in parallel, one for each choice of $(s, k)$, and then disregard all but the one that corresponds to the chosen setting $(k_j, s_j)$. Clearly, this modified algorithm is equivalent to our original algorithm, albeit its inferior computational complexity. The advantage of thinking of our algorithm in this way is that it makes the reduction to the MAB setting very straightforward.

The MAB problem takes place in the space $\mathbb{R}^n$, with $n = nL$. In this space, each coordinate corresponds to a pair $(k, s)$, where $k$ is a cycle length between 1 and $L$, and $s$ is a state in $\mathcal{S}$. For any vector $v \in \mathbb{R}^n$, we use the notation $v_{(k,s)}$ to refer to the element of $v$ that corresponds to the pair $(k, s)$. Let $\rho_{j,(k,s)}$ be the total reward collected by the copy of Alg. 1 that runs with parameters $(k, s)$ on episode $j$. Note that $\rho_{j,(k,s)}$ takes a random value, since Alg. 1 is a randomized algorithm, but this does not invalidate the reduction. The MAB algorithm chooses a pair $(k_j, s_j)$ on each episode, and this is the pair we use in our algorithm.

Next, we prove a regret bound for our algorithm.

**Theorem 2.** *Let $r_1, r_2, \ldots$ be an arbitrary sequence of reward functions, perhaps adversarially generated. Assume that we run MarcoPolo with this sequence for $T$ rounds and with an episode length of $\tau = \sqrt{T}$, and let $(s_t, a_t)_{t=1}^T$ be the resulting sequence of state-action pairs. On the other hand, let $\pi^\star$ be any deterministic policy in $\Pi_{\leq L}$ and let $(s_t^\star, a_t^\star)_{t=1}^T$ be the sequence of state-action pairs defined by $\pi^\star$. Then the regret is upper bounded by*

$$\sum_{t=1}^T r_t(s_t^\star, a_t^\star) - \mathbb{E}\left[\sum_{t=1}^T r_t(s_t, a_t)\right]$$
$$\leq CT^{3/4}\sqrt{\log(T)} + o(T^{3/4}) \ .$$

*where $C = 4L^2(n|\mathcal{A}|)^{3/2} + \sqrt{7nL\log(nL)}$.*

*Proof.* Let $\bar{T}$ be the last round on the last full episode performed by the algorithm. The regret suffered on rounds $(\bar{T}+1), \ldots, T$ is trivially bounded by $T - \bar{T} \leq \tau$, and we focus on bounding the regret on rounds $1, \ldots, \bar{T}$.

Recall that we assumed, for the purpose of our analysis, that we run $nL$ parallel instances of Alg. 1 on each episode. Also recall that Alg. 1 starts from scratch when a new episode begins. Let $k^\star$ be the length of the action cycle induced by the policy $\pi^\star$ and let $\tilde{s}^\star$ be the first state that gets repeated under $\pi^\star$ after an initial non-repeating sequence of states of length $t^\star$. Let $s^\star$ be the state that is $t^\star$ time units behind $\tilde{s}^\star$ on the cycle induced by $\pi^\star$. The rewards accumulated by $\pi^\star$ and those by the cycle corresponding to the pair $(k^\star, s^\star)$ differ by at most $t^\star \leq n$. Thus, we compute regret relative to $(k^\star, s^\star)$.

Let $(s_t', a_t')_{t=1}^{\bar{T}}$ be the sequence of state-action pairs generated by the copy of Alg. 1 that runs with parameters $(k^\star, s^\star)$. On epoch $j$ we apply Thm. 1 and get

$$\mathbb{E}\left[\sum_{t=j\tau-\tau+1}^{j\tau} r_t(s_t^\star, a_t^\star) - r_t(s_t', a_t')\right]$$
$$\leq 4L^2(n|\mathcal{A}|)^{3/2}\sqrt{\tau \log(\tau)} + (2L + d) \ .$$

We sum both sides above for $j = 1, \ldots, \bar{T}/\tau$ and get

$$\mathbb{E}\left[\sum_{t=1}^{\bar{T}} r_t(s_t^\star, a_t^\star) - r_t(s_t', a_t')\right]$$

$$\leq 4L^2(n|\mathcal{A}|)^{3/2}T^{3/4}\sqrt{\log(T)/2} + o(T^{3/4}) \ . \quad (7)$$

Next, we use the regret bound for the EXP3 algorithm, given in Eq. (6), to upper-bound

$$\mathbb{E}\left[\sum_{t=1}^{\bar{T}} r_t(s_t', a_t') - r_t(s_t, a_t)\right]$$
$$= \mathbb{E}\left[\sum_{j=1}^{\bar{T}/\tau} \rho_{j,(k^\star,s^\star)} - \rho_{j,(k_j,s_j)}\right] \ .$$

We face a minor technical difficulty due to the fact that the EXP3 bound holds for deterministic rewards, whereas the rewards $\rho_1, \rho_2, \ldots$ in our problem are random. However, it is rather straightforward to see that if a bound holds for any individual sequence of reward functions, then it also holds for the expected reward sequence (the skeptical reader is referred to Appendix A in the supplementary material for a rigorous proof).

Since $\rho_{j,(k,s)}$ represents the cumulative reward from $\tau$ rounds, and the reward on each round is bounded in $[0, 1]$, then $\|\rho_j\|_\infty \leq \tau \leq \sqrt{T}$. Therefore, we can set $U = \sqrt{T}$ in Eq. (6). Additionally, we have that the number of arms is $nL$. Finally, the number of MAB iterations is $m \leq T/\tau = \sqrt{T}$. Plugging these values into the bound, we get that

$$\mathbb{E}\left[\sum_{t=1}^{\bar{T}} r_t(s_t', a_t') - r_t(s_t, a_t)\right] \leq T^{3/4}\sqrt{7nL\log(nL)} \ .$$

Summing the above inequality with Eq. (7) proves the theorem. □

## 5 DISCUSSION

We focused on the sequential decision making setting where the environment changes with time adversarially, state transitions are deterministic, and the algorithm receives bandit feedback. In this setting, we presented the MarcoPolo algorithm, with an undiscounted regret bound of $O(T^{3/4}\sqrt{\log(T)})$ against the best deterministic policy in hindsight. In contrast to previous work, we did not rely on the stringent unichain assumption.

Many interesting open questions remain unanswered: Can we derive a similar result with stochastic transition dynamics; with adversarially changing transition dynamics; when the state is only partially observable? Even in our specific setting, can we prove a lower bound of $O(T^{3/4})$ on regret, or alternatively, can we obtain a better upper bound with a different algorithm?

We also have more general questions regarding the true power of the adversarial DMDP model compared to the classic (fully stochastic) MDP model. The adversarial DMDP model makes a more general assumption on the rewards and a less general assumption on the state transition dynamics. Are the two things somehow interchangeable? In other words, can we identify situations where we can use the power and flexibility of the adversarial reward assumption to capture uncertainty in the state transition dynamics?

These questions are left for future research.

**Acknowledgements**

**References**

J. Abernethy, E. Hazan, and A. Rakhlin. Competing in the dark: An efficient algorithm for bandit linear optimization. In *Proceedings of the 21st Annual Conference on Learning Theory*, pages 263–274, 2008.

R. Arora, O. Dekel, and A. Tewari. Online bandit learning against an adaptive adversary: from regret to policy regret. In *Proceedings of the 29th International Conference on Machine Learning*, 2012.

P. Auer, N. Cesa-Bianchi, Y. Freund, and R. Schapire. The nonstochastic multiarmed bandit problem. *SIAM Journal on Computing*, 32(1):48–77, 2002.

B. Awerbuch and R. D. Kleinberg. Adaptive routing with end-to-end feedback: Distributed learning and geometric approaches. In *Proceedings of the 36th Annual ACM Symposium on the Theory of Computing*, pages 45–53, 2004.

M. S. Bazaraa, J. J. Jarvis, and H. D. Sherali. *Linear Programming and Network Flows, Third Edition*. John Wiley and Sons, 2010.

D. P. Bertsekas. *Dynamic Programming and Optimal Control*, volume 1. Athena Scientific, Third edition, 2005.

C. Carathéodory. Über den variabilitätsbereich der fourierschen konstanten von positiven harmonischen funktionen. *Rendiconti del Circolo Matematico di Palermo*, 32:193–217, 1911.

D. P. de Farias and N. Megiddo. Combining expert advice in reactive environments. *Journal of the ACM*, 53(5):762–799, 2006.

E. Even-Dar, S. M. Kakade, and Y. Mansour. Online Markov decision processes. *Mathematics of Operations Research*, 34(3):726–736, 2009.

V. F. Farias, C. C. Moallemi, B. Van Roy, and T. Weissman. Universal reinforcement learning. *IEEE Transactions on Information Theory*, 56(5): 2441–2454, 2010.

E. A. Feinberg and F. Yang. On polynomial cases of the unichain classification problem for Markov decision processes. *Operations Research Letters*, 36(5): 527–530, 2008.

A. D. Flaxman, A. Tauman Kalai, and H. B. McMahan. Online convex optimization in the bandit setting: gradient descent without a gradient. In *Proceedings of the 16th annual ACM-SIAM symposium on discrete algorithms*, pages 385–394, 2005.

A. György, T. Linder, G. Lugosi, and G. Ottucsák. The on-line shortest path problem under partial monitoring. *Journal of Machine Learning Research*, 8:2369–2403, 2007.

O. Maillard and R. Munos. Adaptive bandits: Towards the best history-dependent strategy. In *Proceedings of the Fourteenth International Conference on Artificial Intelligence and Statistics*, volume 15 of *JMLR W&CP*, pages 570–578, 2011.

H. B. McMahan and A. Blum. Online geometric optimization in the bandit setting against an adaptive adversary. In *Proceedings of the 17th Annual Conference on Learning Theory*, pages 109–123, 2004.

G. Neu, A. György, C. Szepesvári, and A. Antos. Online Markov decision processes under bandit feedback. In *Advances in Neural Information Processing Systems 23*, pages 1804–1812. MIT Press, 2010.

R. Ortner. Online regret bounds for Markov decision processes with deterministic transitions. *Theoretical Computer Science*, 411(29-30):2684–2695, 2010.

D. Ryabko and M. Hutter. On the possibility of learning in reactive environments with arbitrary dependence. *Theoretical Computer Science*, 405(3):274–284, 2008.

C. Szepesvari. Algorithms for reinforcement learning. *Synthesis Lectures on Artificial Intelligence and Machine Learning*, 4(1), 2010.

J. Y. Yu and S. Mannor. Arbitrarily modulated markov decision processes. In *Proceedings of the IEEE Conference on Decision and Control*, 2009.

J. Y. Yu, S. Mannor, and N. Shimkin. Markov decision processes with arbitrary reward processes. *Mathematics of Operations Research*, 34(3):737–757, 2009.

## A  BANDIT OPTIMIZATION WITH RANDOM REWARDS

The general online bandit optimization problem (with rewards, against an oblivious adversary) is a repeated game between an algorithm and an adversary. Before the game begins, the adversary picks an infinite sequence of reward functions $r_1, r_2, \ldots$ from a function set $\mathcal{F}$. On round $t$ of the game, the algorithm chooses an action $x_t$ from an action set $\mathcal{X}$, the adversary computes the reward $r_t(x_t)$, and reveals this value to the algorithm. The algorithm never gets to see the full function $r_t$.

We define a specific online bandit setting (e.g., the multi-armed bandit setting, bandit linear optimization, etc.) by specifying $\mathcal{X}$ and $\mathcal{F}$. We typically assume that the algorithm's actions are randomized and the adversary's reward functions are deterministic.

**Lemma 3.** *Let A be an online bandit algorithm with a regret bound of $R(T)$ against any **deterministic** sequence of reward functions $r_1, r_2, \ldots$. Namely, for any $T$ it holds that*

$$\max_{x \in \mathcal{X}} \sum_{t=1}^{T} r_t(x) - \mathbb{E}\left[\sum_{t=1}^{T} r_t(x_t)\right] \leq R(T) \;,$$

*where the expectation above is over the algorithm's internal randomization. Then, if each of the reward functions is a **random** function that is independent of the algorithm's randomness, it holds that*

$$\max_{x \in \mathcal{X}} \mathbb{E}\left[\sum_{t=1}^{T} r_t(x) - \sum_{t=1}^{T} r_t(x_t)\right] \leq R(T) \;.$$

*Proof.* Let $\mathscr{F}$ be the $\sigma$-field defined by the random bits used by the random functions $r_1, r_2, \ldots$. In other words, $\mathscr{F}$ is the smallest $\sigma$-field under which all of the random functions are measurable. Loosely speaking, $\mathscr{F}$ captures all of the "randomness" in the reward function and none of the "randomness" used by $A$. When we condition on $\mathscr{F}$, we basically fix a concrete instantiation of the random functions, and for all purposes we make the random functions deterministic. Therefore, it holds that

$$\mathbb{E}\left[\max_{x \in \mathcal{X}} \sum_{t=1}^{T} r_t(x) - \sum_{t=1}^{T} r_t(x_t) \,\Big|\, \mathscr{F}\right] \leq R(T) \;.$$

The left-hand side above is a random variable and the inequality above states that this random variable is surely upper bounded by $R(T)$. We therefore have that

$$
\begin{aligned}
&\max_{x \in \mathcal{X}} \mathbb{E}\left[\sum_{t=1}^{T} r_t(x) - \sum_{t=1}^{T} r_t(x_t)\right] \\
&\leq \mathbb{E}\left[\max_{x \in \mathcal{X}} \sum_{t=1}^{T} r_t(x) - \sum_{t=1}^{T} r_t(x_t)\right] \\
&= \mathbb{E}\left[\mathbb{E}\left[\max_{x \in \mathcal{X}} \sum_{t=1}^{T} r_t(x) - \sum_{t=1}^{T} r_t(x_t) \,\Big|\, \mathscr{F}\right]\right] \\
&\leq R(T) \;.
\end{aligned}
$$

$\square$